# Contamination of TEM Holders Quantified and Mitigated with Open-Hardware, High-Vacuum Bakeout System

Yin Min Goh[1], Jonathan Schwartz[2], Emily Rennich[3], Tao Ma[4], Bobby Kerns[4], Robert Hovden[2,5]

[1] Department of Physics, University of Michigan, Ann Arbor, MI 48109. USA.
[2] Department of Materials Science and Engineering, University of Michigan, Ann Arbor, MI 48109, USA.
[3] Department of Mechanical Engineering, University of Michigan, Ann Arbor, MI 48109, USA.
[4] Michigan Center for Materials Characterization, University of Michigan, Ann Arbor, MI 48109, USA.
[5] Applied Physics Program, University of Michigan, Ann Arbor, MI 48109, USA.

**Corresponding Author:**
Robert Hovden, Ph.D.
H.H. Dow Building, 2300 Hayward St.,
Ann Arbor, MI 48109, USA.
Tel:    770-265-4042
Email: hovden@umich.edu

**Running Title:** TEM Holder Manifold Mitigates Hydrocarbon Contamination

**Abstract**

Hydrocarbon contamination plagues high-resolution and analytical electron microscopy by depositing carbonaceous layers onto surfaces during electron irradiation, which can render carefully prepared specimens useless. Increased specimen thickness degrades resolution with beam broadening alongside loss of contrast. The large inelastic cross-section of carbon hampers accurate atomic species detection. Oxygen and water molecules pose problems of lattice damage by chemically etching the specimen during imaging. These constraints on high-resolution and spectroscopic imaging demand clean, high-vacuum microscopes with dry pumps. Here, we present an open-hardware design of a high-vacuum manifold for transmission electron microscopy (TEM) holders to mitigate hydrocarbon and residual species exposure. We quantitatively show that TEM holders are inherently dirty and introduce a range of unwanted chemical species. Overnight storage in our manifold reduces contaminants by 1 – 2 orders of magnitude and promotes 2 – 4 times faster vacuum recovery. A built-in bakeout system further reduces contaminants partial pressure to below $10^{-10}$ Torr (~4 orders of magnitude down from ambient storage) and alleviates monolayer adsorption during a typical TEM experiment. We determine that bakeout of TEM holder with specimen held therein is the optimal cleaning method. Our high-vacuum manifold design is published with open-source blueprints, parts list, and cost.

**Introduction**

Hydrocarbon contamination degrades resolution and hampers accurate spectroscopic analysis. The formation of insulating carbon films on specimens under electron irradiation has been reported early on and is attributed to the polymerization of organic vapors in a vacuum environment by electric discharge (Stewart, 1934; Love et al., 1981). Hydrocarbon deposition typically results in increased particle size or film thickness accompanied by a loss of contrast (Soong et al., 2012). Increased specimen thickness causes beam broadening which degrades resolution (Watson, 1947; de Jonge et al., 2019). In electron energy loss spectroscopy (EELS), large plasmonic excitations that scale with hydrocarbon thickness (Nerl et al., 2017) combined with scattering from a large carbon-K-edge cross-section become so intense that they obscure core-loss signal from many elements of interest (Fraser, 1978; Griffiths & Walther, 2010; Egerton, 2011).

Besides the notorious hydrocarbons, oxygen and water molecules also pose problems of beam damage and ice contamination in electron microscopy. Oxygen and water molecules absorbed onto specimen surface create highly reactive radicals when irradiated with electrons. These radicals cause lattice damage by chemically etching the specimen—a process sometimes confused with knock-on damage (Leuthner et al., 2019). In cryo-TEM, water surrounding the sample is required to be in a vitreous state. Otherwise, water molecules can form crystalline ice that compromises the structural integrity of a specimen as the crystals withdraw water molecules from the hydration shells. The formation of crystalline ice also degrades image quality as they diffract electrons (Thompson et al., 2016). Hence, it is crucial that the presence of oxygen and water molecules be minimized during the sample preparation and storage phase.

These constraints on high-resolution and spectroscopic chemical imaging demand clean, high-vacuum microscopes with dry pumps. However, even the cleanest microscope columns suffer from impurities desorbed off specimen holders or the specimen itself (Bance et al., 1978); especially problematic in experiments imparting high dose because hydrocarbon deposition scales with beam spot size and current density (Conru & Laberge, 1975). For aberration-corrected scanning transmission electron microscopy (STEM), the high-current density of electrons exacerbates organic polymerization onto specimens. The appearance of contamination can be insidiously delayed, as desorption of species is not immediate and hydrocarbon contamination is driven by surface diffusion of molecules across holder and specimen (Hettler et al., 2017).

Here, we present an open-hardware design of a high-vacuum manifold that stores multiple TEM holders to remedy hydrocarbon and residual species exposure. To confirm the effectiveness of high-vacuum storage, we quantify the molecular species adsorbed onto TEM holder surfaces under various storage conditions using a residual gas analyzer (RGA) as part of our design. Partial pressure measurements by the RGA detect and infer chemical species from their mass-charge ratio (Stanford Research Systems, Inc., 2009) Users can directly assess the composition and cleanliness of holders or specimens. Initial RGA measurements across 7 different TEM holders demonstrate most are inherently dirty and ambient overnight storage will introduce a range of unwanted chemical species into the microscope.

Using overnight storage and bakeout inside our high-vacuum manifold, contaminants partial pressures are reduced by ~4 orders of magnitude to $10^{-10}$ Torr (below the RGA detection limit). Overnight high-vacuum storage reduces residual gas levels across the whole spectrum by 1 – 2 orders of magnitude (~$10^{-7}$ Torr) and promotes 2 – 4 times faster vacuum recovery. Integration of a bakeout system is substantially effective at removing problematic pump oils and reducing atmospheric species by an additional 2 – 3 orders of magnitude down to below $10^{-10}$ Torr. Reducing contaminants partial pressure to below $10^{-10}$ Torr alleviates monolayer adsorption during a typical TEM experiment. Hence, we determine that high-vacuum storage and thermal bakeout of holders with specimens held therein is the optimal storage and regulation method for high-resolution electron microscopy.

Our manifold design consists of a 2-tier structure that stores up to 10 TEM holders and can be easily customized to suit a facility's needs. Open-source blueprints, part lists, and costs are provided for all electron microscope facilities. The design is inspired by manifolds previously built at Cornell University and McMaster University. Nion Co. uses high-vacuum bakeout routines for their cartridge specimen stages (Krivanek et al., 2008). However, only a handful TEM facilities store holders in high vacuum.

## Results

**The Contaminants Present on TEM Holders**

The RGA spectrum for a typical TEM holder (Fig. 1a) highlights the range of species adsorbed onto holder surfaces, spanning organics of various carbon compositions, viscous pump oil, water and oxygen, which totals a manifold pressure of $10^{-4}$ – $10^{-5}$ Torr (hPa). Initial RGA measurements taken across six other TEM holders regularly used and stored in ambient room conditions (Fig. 1b) demonstrate that most are inherently dirty and introduce a range of unwanted chemical species into vacuum. Several regions of the RGA spectrum contain chemical species that commonly degrade electron micrographs (Jenninger & Chiggiato, 2017). These contaminants originate from atmospheric organics, microscope pump oil, and o-ring vacuum greases – all of which can accumulate on TEM holder surfaces without proper storage and regulation.

Vacuum systems using oil-based rotary and diffusion pumps are often considered "dirty" systems because they produce backstreaming of oil (Postek, 1996). The oil vapor can polymerize under the electron beam, resulting in the deposition of amorphous carbon on the area of investigation (Ennos, 1953). Oil-free systems, such as scroll, turbomolecular and ion pumps should replace oil-based pumps for evacuating S/TEM columns (Mitchell, 2015). In addition to improvements in pumps, contamination within the column can be mitigated with a cold trap (Ennos, 1953), also known as an anti-contamination device (ACD). The ACD is a liquid nitrogen cooled trap that condenses vapors near the specimen to minimize their redeposition (Yoshimura et al., 1983). It should be mentioned that any other cold surface in the system will act the same way (e.g. x-ray detectors) (Reimer & Wächter, 1978).

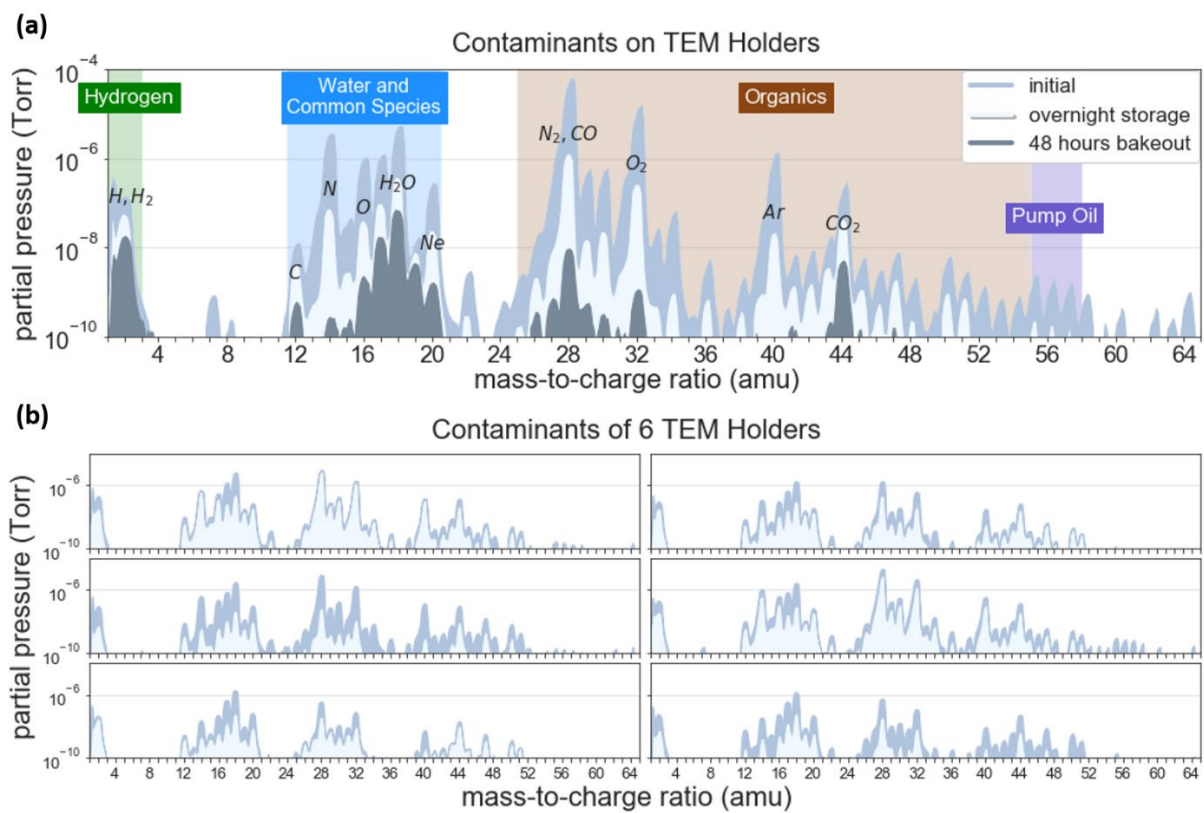

**Figure 1.** (**a**) RGA spectrum of a dirty JEOL TEM holder cleaned from overnight high-vacuum storage and bakeout at 130 ºC for 48 hours. Shaded regions highlight chemical species that commonly degrade electron micrographs. Specifically, oxygen and water (16-18 amu), organics molecules (25-55 amu), and pump oil hydrocarbons (55-59 amu) are typical regions of concern to microscopists. High-vacuum storage lowers the residual gas levels accumulated from ambient conditions by 1 – 2 orders of magnitude. Bakeout further reduces organic and light species by another 2 – 3 orders of magnitude. Heavier species above 35 amu including pump oil are mostly below the detectable limit (< $10^{-10}$ Torr). (**b**) RGA spectra of six other TEM holders regularly used and stored under ambient conditions with partial pressure lowered after overnight storage in the manifold.

Side entry TEM holders also use long-term high-vacuum greases based on perfluorinated polyether (PFPE) oils and polytetrafluoroethylene (PTFE) thickeners that out-gas and contaminate specimens. X-ray photoelectron spectroscopy (XPS) analysis of a TEM specimen grid that has gone through 10 pumping cycles in a modern TEM equipped with ACD shows emerging fluorine peaks compared to a clean grid that was never loaded into a TEM (Supplemental Fig. S1). However, XPS does not show a significant increase in carbon signals. Vacuum grease should be used in minimal quantities. Vacuum grease may also accumulate in the loadlock or goniometer of electron microscopes and should be cleaned semi-regularly. For TEM o-rings, micronized PFPE grease (e.g. Y VAC, Braycote Micronic 1613, Krytox) is preferred. Avoid UHV hydrocarbon-based grease (Rheolube) that releases hydrocarbon byproducts, pure Fomblin with viscosity issues, or clear silicon-based high vacuum grease (Dow Corning) with relatively high vapor pressure ($10^{-6}$ Torr). As aforementioned, diffusion pump oil is a source of contamination and must not be used on o-rings.

**Clean Holders with High-Vacuum Storage and Bakeout**

Our high-vacuum manifold effectively reduces contaminant partial pressures by ~4 orders of magnitude to even below $10^{-10}$ Torr when used for overnight storage and bakeout (Fig. 1a). With overnight high-vacuum storage alone, residual gas levels across the whole spectrum reduce by 1 – 2 orders of magnitude (~$10^{-7}$ Torr). A built-in bakeout system is substantially effective at removing problematic pump oils and reducing atmospheric species (i.e. CO, $H_2O$, etc.) by an additional 2 – 3 orders of magnitude down from high-vacuum storage and ~4 orders lower compared to storage in ambient air. The adsorption coverage described by the Langmuir isotherm is a worst case scenario (sticking coefficient = 1) that will form one monolayer per second at pressures of $10^{-6}$ Torr (Jousten, 1999); reducing partial pressures of contaminants below $10^{-10}$ Torr (the RGA detection limit) slows the monolayer adsorption time to ~5 hours. We consider partial pressures below $10^{-10}$ Torr negligible.

By facilitating organic desorption through bakeout, the presence of light species is reduced by several orders of magnitude, while pump oil and most of the heavier species above 35 amu were below the RGA's detectable limit (< $10^{-10}$ Torr) (Fig. 1a). Gas molecules on a surface can be described as a distribution of binding energies. The thermal outgassing rate for a gas species is an exponential decay function of $E/k_BT$ and depends on initial surface coverage. Molecules across all binding energy states are more likely to desorb with increasing bakeout temperature. Even with vacuum bakeout limits (150 – 250ºC) well below the average binding energies (0.73 – 1.08 eV/molecule), molecules in low binding energy states can slowly desorb over long pumping time (Matthewson & Gröbner, 1999) and minimize migration driven by surface diffusion of organic molecules that do not immediately desorb in vacuum (Dayton, 1961; Hettler et al., 2017). Following Boltzmann statistics, baking at 130ºC increases desorption rate by 35% over room temperature. Our design utilizes radiative heating from a quartz lamp that operates directly above the holder inside the manifold (Fig. 5). This method allows for radiative heat energy to reach the holder in vacuum while also preventing the polymer o-rings from overheating as shown in Fig. 5a. After baking, the manifold achieves its lowest total pressure of ~$7.5 \times 10^{-8}$ Torr ($10^{-7}$ hPa) even when a holder is stored.

We determine that thermal bakeout in vacuum exhibits higher performance over chemical cleaning, that leaves organic residues and plasma cleaning that may damage carbon-containing specimens and only removes thin layers of surface hydrocarbons. Cleaning the holder with solvents (i.e. acetone and/or methanol) can remove the majority of organic contamination, but commonly leaves organic residue (McGilvery et al., 2012). Fortunately, overnight manifold storage will remove any organic residue introduced from a chemical clean. Supplemental Figure S2 shows chemical cleaning with alcohols introduces organic peaks in residual gas levels resembling that of acetone. These alcohols will desorb after 6 hours storage in high vacuum. Plasma cleaning can immobilize and remove thin layers of surface hydrocarbons, which is sufficient for superficial contamination. The holder in Figure S2 was subsequently plasma cleaned for 10 minutes on a 40 W rated plasma cleaner operating at the standard RF-plasma frequency (13.56 MHz) with a mixture of oxygen (25%) and argon (75%). The holder is immediately transferred to the manifold stationed on the same lab bench. RGA data shows that carbon level remains relatively unchanged and that only peaks ~18 amu increase after plasma cleaning. This could be moisture adsorbed onto the surface during the transfer from plasma cleaner to manifold.

The oxygen plasma chemically reacts with hydrocarbons and converts it into CO, $CO_2$, and $H_2O$, which are subsequently evacuated by the vacuum system (Isabell et al., 1999). One major limitation with plasma cleaning is its potential to damage any carbon support films or carbon containing specimens that may reside on the holder. Our design can address this issue by removing contaminants while preserving any samples capable of withstanding moderate heating.

Bakeout of TEM holders and the specimens held therein, are carried out inside the manifold at 130 ºC for 48 hours as organic molecules desorb at this temperature (Jousten, 1998; Grinham & Chew, 2017) without degrading o-rings (Viton ~225ºC, Buna ~120ºC) and internal wiring components of a TEM holder. Calibration of bakeout temperature is based on thermal readings with a Gatan #652 heating holder made of beryllium copper and tantalum. Bakeout is also tested on JEOL single- and double-tilt holders. In all these cases vacuum heating is equally capped at 130 °C and reached $10^{-7}$ hPa total pressure. While (austenitic) stainless steel and aluminum alloys make up many vacuum components, certain grades (i.e. aluminum alloy 7000 series, stainless steel free-machining grades) and some metals such as brass, cadmium, zinc, tellurium and Pb-based solder flux are not vacuum compatible in the first place (Coyne, 2013). In-situ gas or liquid cell holders may contain polymer tubing (e.g. PEEK) that cannot be baked above 100°C. We opted for Viton o-rings due to its high thermal stability and low outgassing and permeability. Small outgassing and permeation rates are essential to reach low base pressures. Lubrication of o-rings in minimal quantities (one drip for the entire o-ring) is necessary to protect it from abrasion and degradation damages by atmosphere. We use Klüberalfa Y VAC O-ring grease due to its high thermal stability and low vapor pressure. We recommend annual replacement of o-rings and vacuum grease or when o-rings have experienced frequent heat cycling.

**High-Vacuum Storage for Faster Pump down**

A TEM holder exposed to ambient air for 10 minutes (roughly the time to load a specimen) after high-vacuum storage achieves partial pressure recovery 2 – 4 times faster than that of a holder stored in ambient air. Figure 2 compares the manifold total pressure of TEM holders exposed to ambient air for over 1 day—as found in a typical TEM facility—and those stored in high-vacuum with only 10 minutes of ambient exposure. These pressures are measured at the top of the manifold vacuum column. We define full recovery to be the minimum total pressure a manifold with a stored TEM holder can achieve that remains constant thereafter for at least one day. Partial recovery is defined to be the $4.0\times10^{-7}$ hPa mark in Fig. 2 and lower pressures. In our comparison, full recovery takes 6 hours and 3.5 hours respectively, while partial recovery takes 2 hours and 30 minutes respectively. Thus, partial recovery is 4 times faster and full recovery reaches nearly two times lower pressure in half the time.

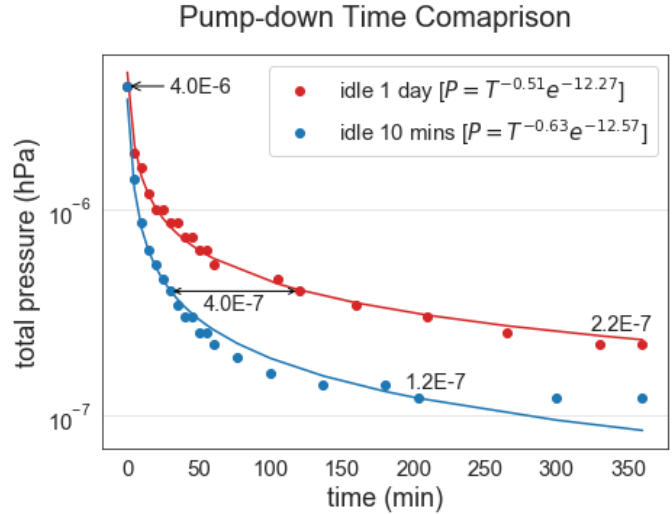

**Figure 2.** TEM holder stored in high vacuum (blue) achieves faster pump down and reaches lower base pressure compared to the one stored in atmosphere (red). The fitted model is non-physical but explicitly compares the total pressure decay rate.

Given that mounting a sample takes around 10 minutes, these results suggest users can achieve faster pressure recovery within the TEM column and have more efficient microscopy sessions with high-vacuum storage. A typical TEM column requires evacuation down to $10^{-7}$ Torr (hPa) by turbomolecular (dry) or diffusion (oil) pumps operating in the high vacuum range [12]. When a TEM holder is clean, the manifold can achieve a total pressure as low as $\sim 7.5 \times 10^{-8}$ Torr ($10^{-7}$ hPa), which is the usual limit of o-ring sealed systems. The improved vacuum and pumping speeds reflect the overall reduction of adsorbed species for TEM specimen holders stored under high vacuum.

## Materials and Methods
### Open-Hardware Design of a TEM Holder Manifold

Our design consists of a 2-tier structure that can store up to 10 TEM holders. The number of tiers and ports can be easily customized to suit each facility's needs. In our case, we opted for a 2-tier structure to store both our FEI and JEOL holders. The whole manifold is assembled on top of a Pfeiffer Vacuum HiCube 80 Eco turbopump station supported by an 80/20 extruded aluminum frame (Fig. 3a). The Pfeiffer pumping station includes a control unit for the turbopump and a diaphragm backing pump. Pump down and ventilation are controlled using the display unit, from which turbo speed, frequency, and total pressure of the system can be inspected. Pressure and mass measurements are collected with a Pfeiffer Vacuum Pirani/cold cathode (PKR) pressure gauge and SRS RGA (Fig. 3c).

The RGA sorts and detects the ion current of residual gases based on their mass-to-charge ratio. We have opted for the 100 amu model, which allows for identification of the common contaminants found in electron microscopy applications (gas species ranging from 1 - 100 amu). The RGA is installed on the manifold backside for structural stability and to provide sufficient space for the quadrupole mass probe. The RGA also includes an electronic control unit, cable wires for computer connection while avoiding interruption during holder insertion and removal. Mass spectra using the RGA are acquired from an adjacent computer.

Each tier contains an array of butterfly valves that opens each holder to vacuum. The pressure measurements in Figure 2 were taken with several ports solely sealed by butterfly valves indicating that these valves are sufficient at isolating the vacuum against atmosphere at or below that of the TEM o-ring seal limit (~ $10^{-7}$ hPa). Scientists may also substitute more expensive gate valves that can handle higher pressure differentials, or all metal angle valves that can tolerate higher temperature bakeouts. However, total pressure lower than $10^{-7}$ hPa are not easily achieved in o-ring sealed vacuum system. Due to the unique diameters, the pipes were custom made by MDC Vacuum Products. The designs for custom pipes to fit a JEOL TEM holder are shown in Supplemental Figure S3. The bakeout system consists of a quartz lamp to an electrical feedthrough installed in the mini side port of storage flanges (Fig. 5). This allows for radiative heating of the TEM holder tip and specimen (if mounted) up to ~150 ºC while in vacuum. Bakeout temperature is variable with applied electric potential. To achieve higher vacuum, we opted for ConFlat (CF) flanges over ISO-KF / ISO-LF.

Operation of the manifold is straightforward and requires only a few minutes of training. Holder exchange requires shutting off the turbopump and complete ventilation of the chamber to avoid damaging the turbo blades. The turbo will audibly wind down within 5 minutes as the pumping station self-vents. The display panel can be used to inspect turbopump speeds to ensure the blades are not rotating before exchanging the holder. The butterfly valve of the holder flange in exchange needs to be shut off to keep the rest of chamber in low vacuum ($10^{-1}$ -$10^{-3}$ hPa). Some resistance may be felt when pulling the holder out due to higher pressure outside the manifold. A more cavalier approach of closing the butterfly valve and removing the holder against vacuum can be done without winding down the turbo. Before turning the turbopump back on, the butterfly valves for all unused ports must be closed. Dummy holders may be optionally inserted into unused ports for additional safety. Dummy holders are easily machined from the included plans (Supplemental Fig. S4).

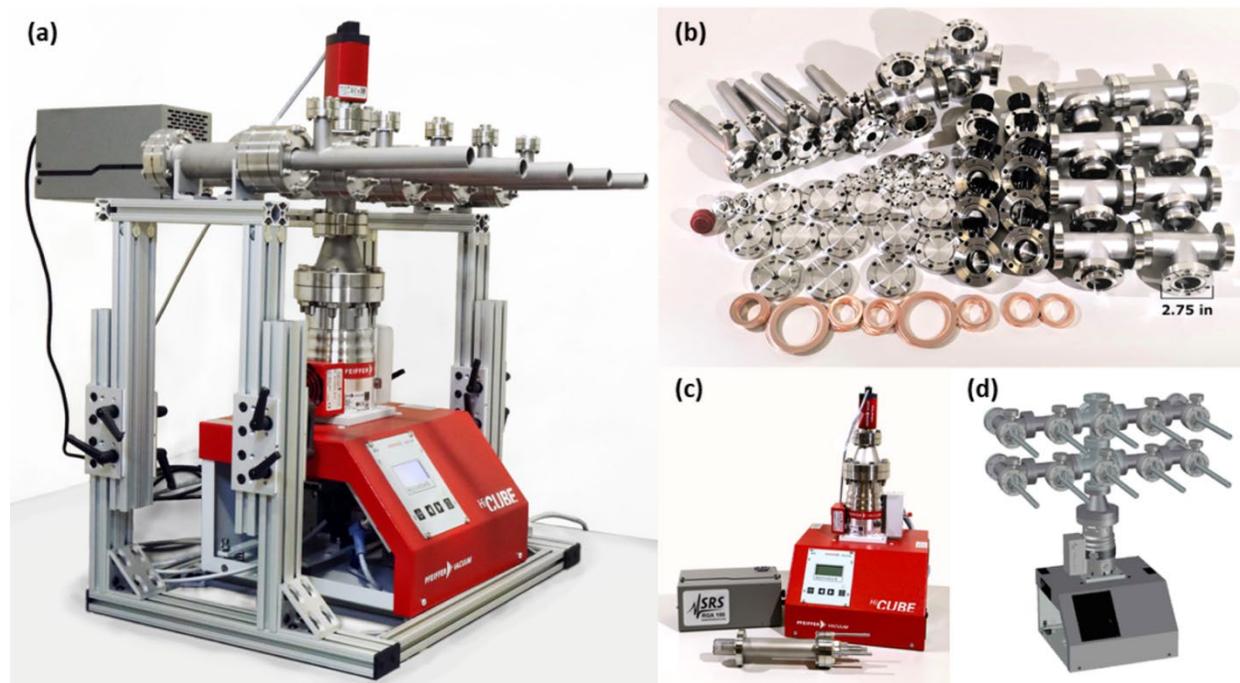

**Figure 3.** Components and complete set-up of the vacuum manifold to store 10 TEM holders. (**a**) Operating high-vacuum manifold on a support frame at a TEM facility; our design is customizable up to 2 tiers with 10 ports. (**b**) Vacuum compatible stainless-steel parts and copper gaskets. (**c**) Stanford Research Systems 100 amu Residual Gas Analyzer (RGA) with quadrupole mass probe, Pfeiffer Vacuum HiCube 80 Eco turbopump station with Pfeiffer Vacuum ActiveLine PKR total pressure gauge. (**d**) CAD $_{drawing}$ of manifold assembly. The design is made open-source.

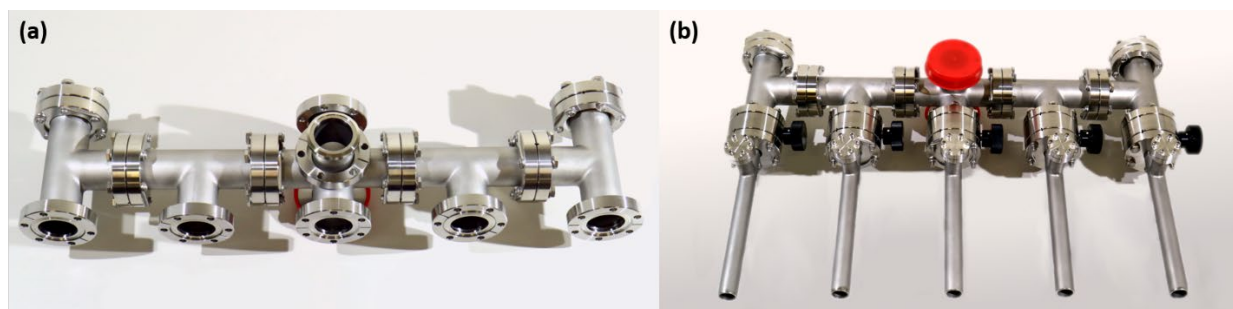

**Figure 4.** Assembly plan for one tier of manifold. (**a**) Configuration of tees with respect to six-way cross. (**b**) Butterfly valves and pipe flanges connection on one side of the tees opening.

**Parts and Assembly**

Assembly is completed in multiple stages starting with the first tier. It is recommended to work with nitrile gloves, maintain clean working surfaces, and wrap components with aluminum foil to minimize atmospheric exposure. Each tier consists of a six-way cross connected to 4 tees (Fig. 4a). Extra or unused flanges should be blanked. The butterfly valve and custom pipe are tightened to the tees by 2" hex bolts. For uniform orientation, the hand wheels on the valves should all be pointing right (Fig. 4b). After assembling the tier, it will be attached onto the pumping station cross. A conical reducer is placed between the 4.5" turbopump flange and 2.75" six-way cross flange (Fig. 3d). The PKR gauge and RGA are installed on the top and back of six-way cross, respectively (Fig. 3a). The ionizer cage side of RGA should go inside six-way cross, while the probes point behind and align with the holes on electronic control unit until complete contact is achieved. When connecting vacuum components, a fresh copper gasket is first placed against the knife edge seal of a CF flange. At the mating flange, bolts are hand tightened with small increments in a crisscross star sequence to prevent over-tightening on one side and a bad seal.

The Pfeiffer HiCube 80 model has a maximum rotation speed of 90,000 rpm and a frequency of 1,500Hz. When these parameters cannot not reach the maximum values, a leak may be occurring at the vacuum connections and the system should be partially disassembled and pumped down to search for leaks. If a CF flange is reopened, any used copper gasket must be replaced. Once the chamber is fully assembled and tested for leaks, the system is left to pump down overnight to reach an initial base pressure. To achieve lower pressure and remove any remaining adsorbed molecules, the manifold is baked overnight at 130°C with a BriskHeat silicone rubber heating tape. Note that the heating tape can damage the butterfly valves if wound too closely around them. An initial pump down is critical when installing the RGA as it cannot operate at pressures above $10^{-4}$ Torr. The RGA electronic control unit must be removed prior to this process, but bake-out of the RGA mass probe is still recommended.

Setting up the bake out chambers requires the following steps. A bi-pin type quartz lamp and power supply wiring are soldered to the Type-C 9 pins subminiature electrical feedthrough. Lead and halogen-free solder is used to avoid outgassing in vacuum. The soldered quartz lamp is cleaned with a flux remover and organic solvents (e.g. acetone and methanol) in an ultrasonicator. The lamp is fitted to 2 pins on the vacuum side of electrical feedthrough and assembled on the mini port with the lamp body inside manifold (Fig. 5c). To begin bakeout, power supply wiring is connected to the electrical feedthrough (Fig. 5a). Our calibration using a heating holder determined that bake-out temperatures of ~130°C can be achieved with 39 - 40V at a constant current of 0.22A from a 120V, 50W rated quartz lamp.

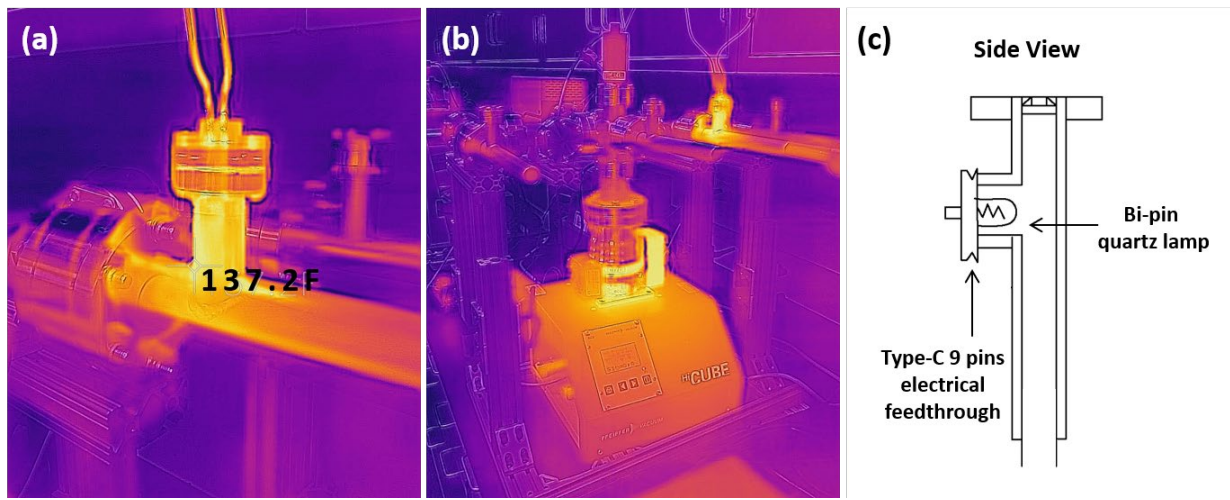

**Figure 5.** Radiative heating of TEM holder in high-vacuum. Our bakeout system consists of a quartz lamp to an electrical feedthrough with the bakeout temperature variable by applied voltage. Thermal images taken on a forward-looking infrared (FLIR) camera show that **(a)** the maximum external temperature is ~137°F (58°C) with o-ring sealed butterfly valve relatively cool. **(b)** The turbopump and bakeout flange are at high temperature while the rest of the manifold is safe for contact during a bakeout. **(c)** Schematic of custom pipe flange shows the mini side port where the bakeout system is installed.

Table 1: Components and Costs for Manifold

| Part Details | Cost |
|---|---|
| HiCube 80 Eco DN 63 CF-F & ActiveLine PKR DN 40 CF-F | $6,000 |
| RGA 100 Residual Gas Analyzer | $3,750.00 |
| 80/20 Mainfold Support Frame | $944.41 |
| Satco T4 Bi-Pin Quartz Lamp **(2)** | |
| B&K Precision 1685B DC Power Supply | $371.00 |
| Kester 268 Flux-Cored Wire (Sn96.5Ag3.0Cu0.5 3.3%/268 .020 500 G Robo SPL) | $89.79 |
| Molex EconoLatch Female Terminal 20-22 AWG Pin Connector **(25)** | $0.039 |
| Chemtronics ES132 Flux-Off Aqueous | $39.41 |
| Klüberalfa Y VAC Vacuum Grease | $127.00 |
| 6-Way Cross, 2.75" Tube **(2)** | $612.00 |
| Tee, 2.75" **(8)** | $128.00 |
| Butterfly Valve, 2.75" OD **(10)** | $287.00 |
| Flange, 1.33" **(10)** | $15.00 |
| Flange, 2.75" **(10)** | $17.00 |
| Copper Gaskets, 1.33" **(3)** | $17.00 |
| Copper Gaskets, 2.75" **(4)** | $22.00 |
| Hex Head Bolt/Plate Nut, 1.25" **(4)** | $31.00 |
| Female, 25"NPT x 2.75" Flange **(2)** | $97.00 |
| Socket Head Screw/Plate Nut, .75" **(3)** | $27.00 |
| Conical Reducer, 4.5" x 2.75" | $153.00 |
| Copper Gasket, 4.5" Flange | $38.00 |
| Hex Head Bolt/ Plate Nut, 2" Lg | $58.00 |
| Type-C 9 Pins Del Seal CF **(2)** | $274.00 |
| Custom Flange with Mini CF Side Port **(10)** | $559.00 |
| **TOTAL** | **$23,570.57** |

**Conclusion**

Hydrocarbon contamination has been an ongoing challenge for microscopists as it plagues image resolution, contrast, and chemical sensitivity. Here we quantitatively show that TEM specimen holders are a notable source for hydrocarbon contamination and water vapor during an electron microscope experiment. To significantly mitigate this problem, we constructed a high-vacuum TEM holder manifold that provides roughly a two-order magnitude reduction in water vapor and common organic species. Our TEM holder manifold design is published as an open-hardware project with a parts list and design plans. In the future, we hope to see the development of higher vacuum inside the microscope column and routine implementation of dry systems such as turbomolecular or ion pumps.

**License:**

**Acknowledgements:** Authors recognize support and inspiration from other facilities with custom in-house TEM manifolds—Cornell University (Malcolm Thomas, Lena F. Kourkoutis, John Grazul) and the Canadian Centre for Electron Microscopy at McMaster University (Andy Duft, Glynis de Silviera, Gianluigi Botton). We thank Eckhart / Autocraft for the frame support design. R.H. and Y.M.G. received support from the National Science Foundation (NSF #1807984).

**References:**

Stewart, R.L. (1934). Insulating Films Formed Under Electron and Ion Bombardment, *Phys. Rev.* **45**, 488–490. https://doi.org/10.1103/PhysRev.45.488.

Love, G., Scott, V.D., Dennis, N.M.T. & Laurenson, L. (1981). Sources of contamination in electron optical equipment. *Scanning* **4**, 32–39. https://doi.org/10.1002/sca.4950040105.

Soong, C., Woo, P., Hoyle & D. (2012). Contamination Cleaning of TEM/SEM Samples with the ZONE Cleaner. *Microsc. Today* **20**, 44–48. https://doi.org/10.1017/S1551929512000752.

Watson, J.H.L. (1947). An Effect of Electron Bombardment upon Carbon Black. *J. Appl. Phys.* **18**, 153–161. https://doi.org/10.1063/1.1697597.

de Jonge, N., Houben, L., Dunin-Borkowski, R.E. & Ross, F.M. (2019), Resolution and aberration correction in liquid cell transmission electron microscopy. *Nat. Rev. Mater.* **4**, 61–78. https://doi.org/10.1038/s41578-018-0071-2.

Nerl, H.C., Winther, K.T., Hage, F.S., Thygesen, K.S., Houben, L., Backes, C., Coleman, J.N., Ramasse, Q.M. & Nicolosi, V. (2017). Probing the local nature of excitons and plasmons in few-layer $MoS_2$. *npj 2D Mater. Appl.* **1**, 2. https://doi-org.proxy.lib.umich.edu/10.1038/s41699-017-0003-9.

Fraser, H.L. (1978). Elemental analysis of second-phase carbides using electron energy-loss spectroscopy. In *Scanning Electron Microscopy*, Johari, O. (Ed.), Part 1, pp. 627–632. SEM Inc., A. M. F. O'Hare, Chicago, IL.

Griffiths, A.J.V. & Walther, T. (2010). Quantification of carbon contamination under electron beam irradiation in a scanning transmission electron microscope and its suppression by plasma cleaning. *J. Phys.: Conf. Ser.* **241**, 012017. https://doi.org/10.1088/1742-6596/241/1/012017.

Egerton, R.F. (2011). *Electron Energy-Loss Spectroscopy in the Electron Microscope.* US: Springer, Boston, MA. https://doi.org/10.1007/978-1-4419-9583-4.

Leuthner, G.T., Hummel, S., Mangler, C., Pennycook, T.J., Susi, T., Meyer, J.C. & Kotakoski, J. (2019). Scanning transmission electron microscopy under controlled low-pressure atmospheres. *Ultramicroscopy* **203**, 76–81. https://doi.org/10.1016/j.ultramic.2019.02.002.

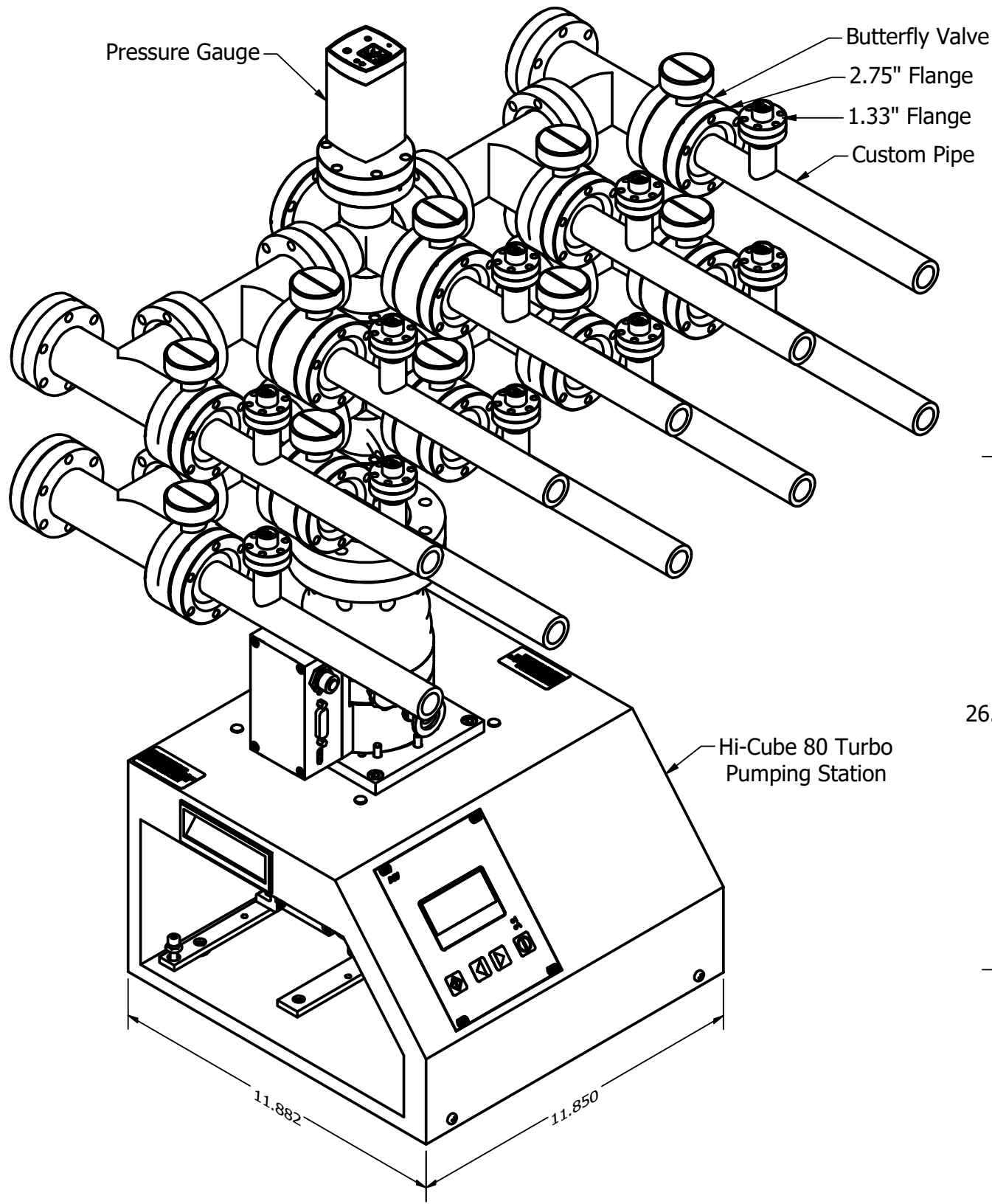
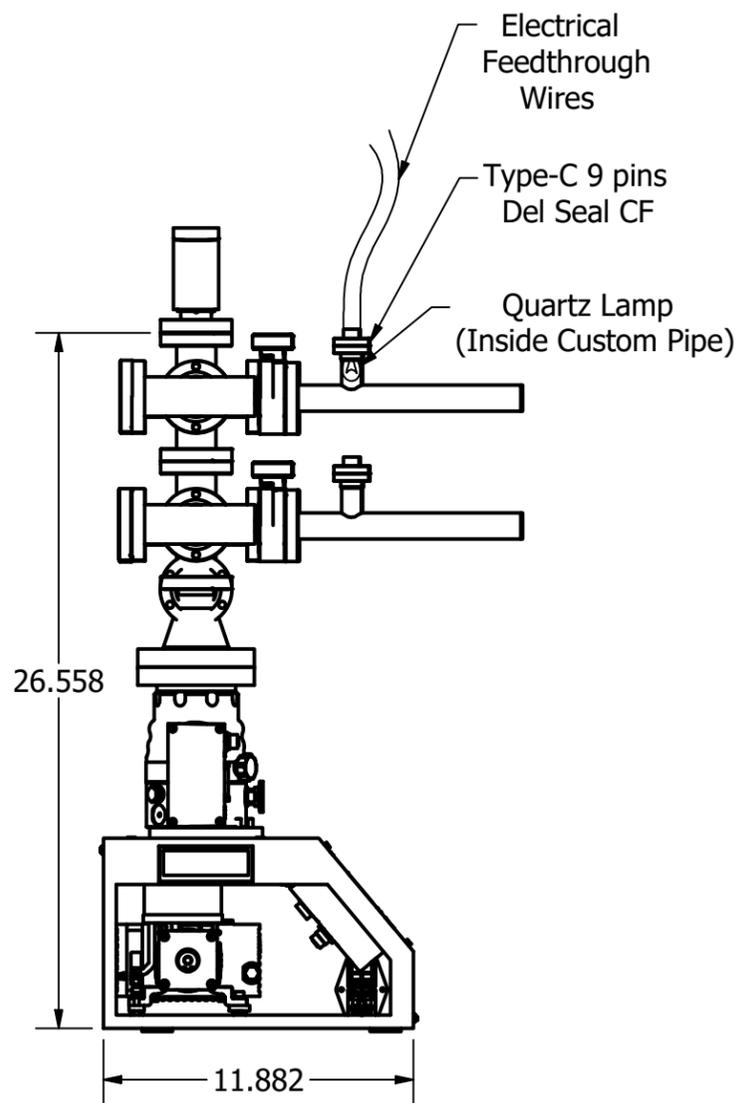
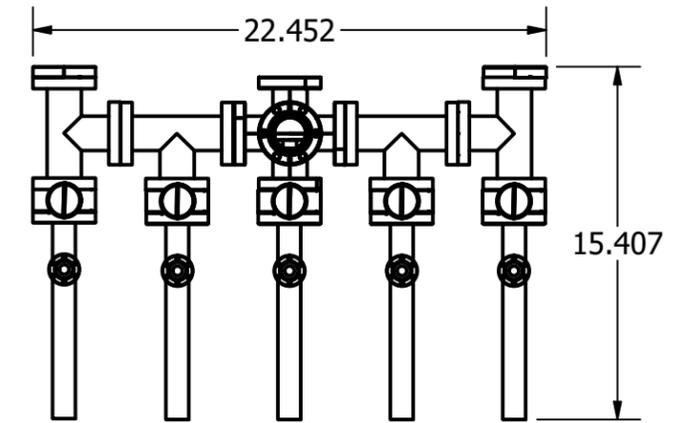

| QTY | MANIFOLD PARTS LIST<br>PART |
|---|---|
| 1 | HiCube 80 Eco DN 63 CF-F & ActiveLine PKR DN 40 CF-F |
| 1 | RGA 100 Residual Gas Analyser |
| 1 | 80/20 Manifold Support Frame |
| 2 | Satco T4 Bi-Pin Quartz Lamp |
| 1 | B&K Precision 1685B DC Power Supply |
| 1 | Kester 268 Flux-Cored Wire |
| 25 | Molex EconoLatch Female Terminator 20-22 AWG Pin Connector |
| 1 | Chemtronics ES132 Flux-Off Aqueous |
| 1 | Kluberalfa Y VAC Vacuum Grease |
| 2 | 6-Way Cross, 2.75" Tube |
| 8 | Tee, 2.75" |
| 10 | Butterfly Valve, 2.75" OD |
| 10 | Flange, 1.33" |
| 10 | Flange, 2.75" |
| 3 | Copper Gaskets, 1.33" (packs of 10) |
| 4 | Copper Gaskets, 2.75" (packs of 10) |
| 4 | Hex Head Bolt/Plate Nut, 1.25" (packs of 24) |
| 2 | Female, 25" NPTx2.75" Flange |
| 3 | Socket Head Screw/Plate Nut, 0.75" |
| 1 | Conical Reducer, 4.5"x2.75" |
| 1 | Copper Gasket, 4.5" Flange |
| 1 | Hex Head Bolt/Plate Nut, 2" Lg (packs of 24) |
| 2 | Type-C 9 Pins Del Seal CF |
| 10 | Custom Flange with Mini CF Side Port |
| 1 | P11 O-rings for JEOL dummy holder (packs of 10) |
| 1 | FKM (Viton) 2x8.5 O-rings for FEI dummy holder (packs of 10) |

**Contamination of TEM Holders Quantified and Mitigated with Open-Hardware, High-Vacuum Bakeout System**

Yin Min Goh, Jonathan Schwartz, Emily Rennich, Tao Ma, Bobby Kerns, Robert Hovden

Journal: Microscopy and Microanalysis (In Review)

Volume | Issue | Page Number | Date of Issue

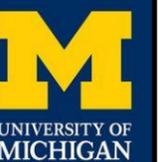

*Note: Manifold can be coupled with a support frame for additional sturdiness

This open hardware design is released under the CERN-OHL-W version 2 license

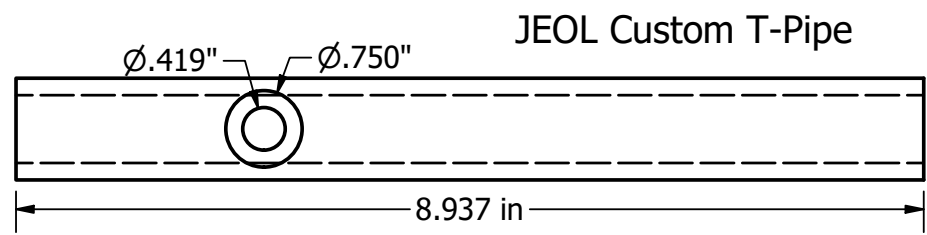

JEOL Custom T-Pipe
Ø.419"  Ø.750"
8.937 in
1.669 in
Ø1.000" / Ø.669"
1.500"

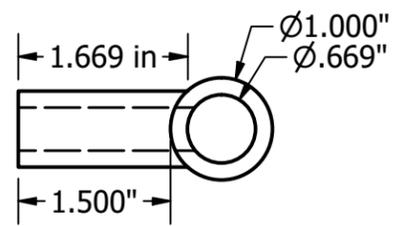

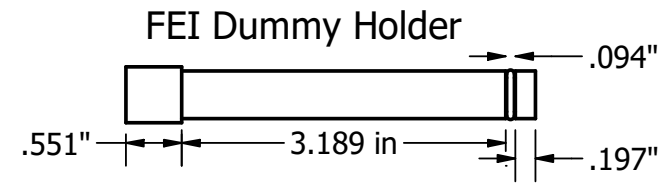

FEI Dummy Holder
.551"  3.189 in  .094" .197"

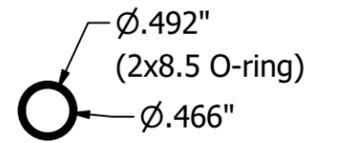

Ø.492" (2x8.5 O-ring)
Ø.466"

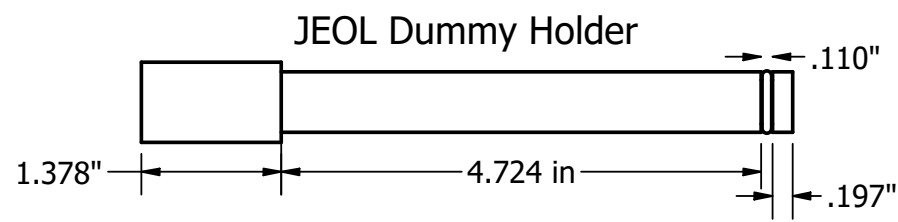

JEOL Dummy Holder
1.378"  4.724 in  .110"  .197"

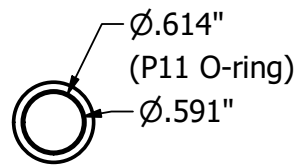

Ø.614" (P11 O-ring)
Ø.591"

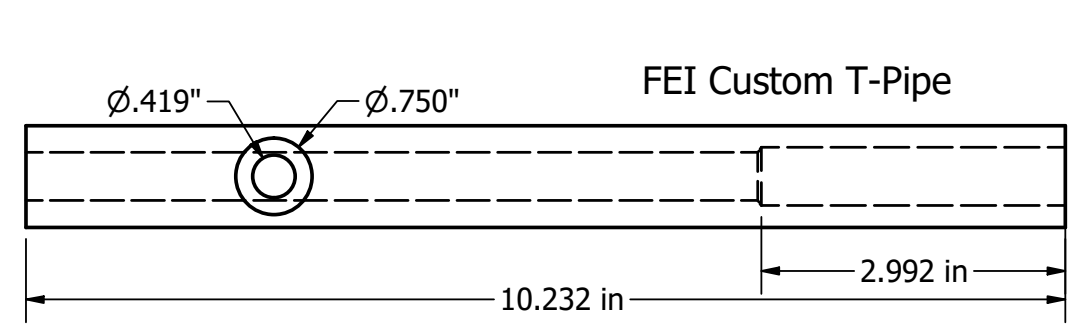

FEI Custom T-Pipe
Ø.419"  Ø.750"
10.232 in  2.992 in
1.669 in  Ø1.000" / Ø.571" / Ø.472"
1.500"

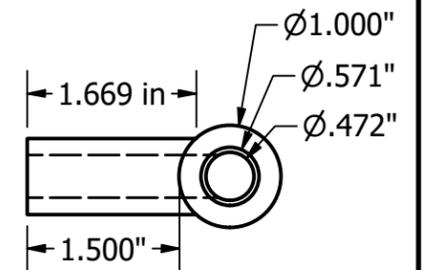

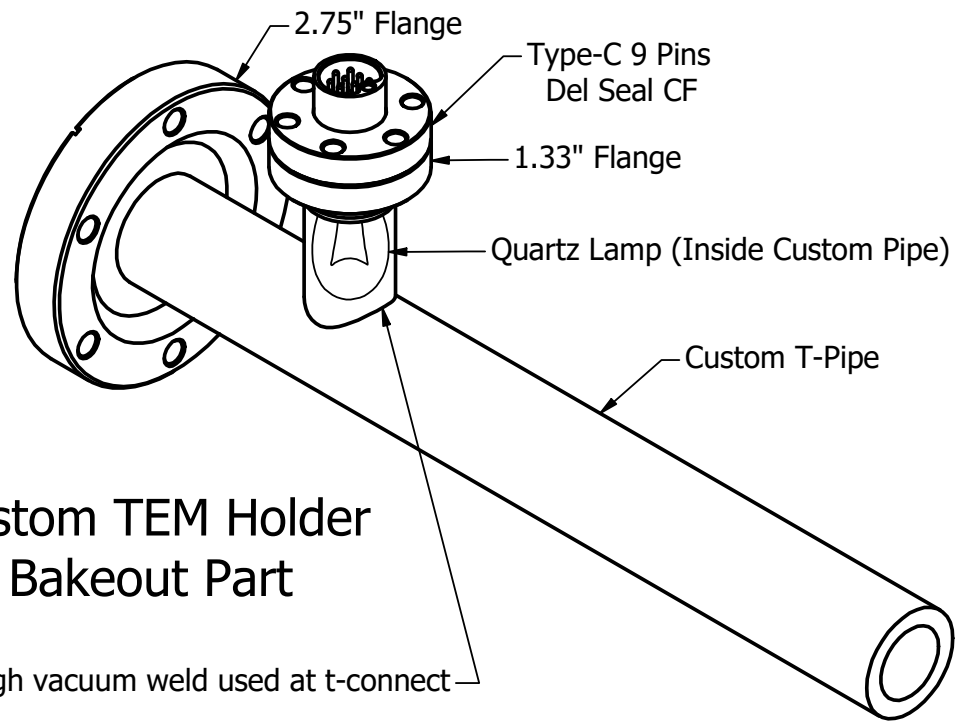

Custom TEM Holder Bakeout Part

2.75" Flange
Type-C 9 Pins Del Seal CF
1.33" Flange
Quartz Lamp (Inside Custom Pipe)
Custom T-Pipe
High vacuum weld used at t-connect

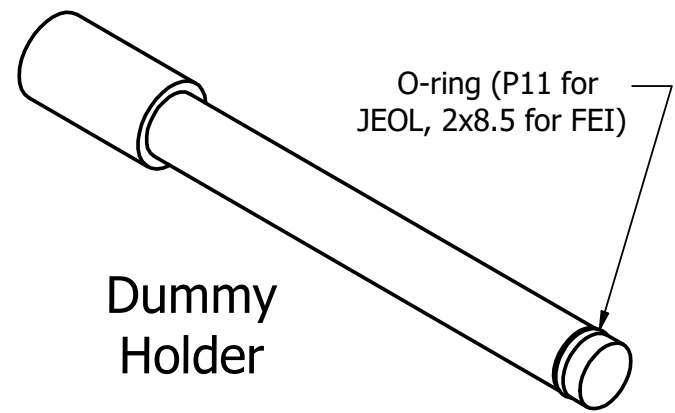

Dummy Holder

O-ring (P11 for JEOL, 2x8.5 for FEI)

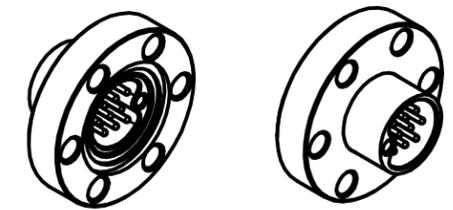

Type-C 9 Pins Del Seal CF

**Contamination of TEM Holders Quantified and Mitigated with Open-Hardware, High-Vacuum Bakeout System**

Yin Min Goh, Jonathan Schwartz, Emily Rennich, Tao Ma, Bobby Kerns, Robert Hovden

Journal: Microscopy and Microanalysis (In Review)

| Volume | Issue | Page Number | Date of Issue |

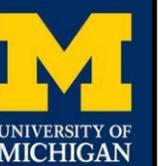